# Evaluating discriminatory accuracy of models using partial risk-scores in two-phase studies

PARICHOY PAL CHOUDHURY

*Department of Biostatistics, The Johns Hopkins University, Baltimore, MD, USA*

parichoy@jhu.edu

ANIL K. CHATURVEDI

*Division of Cancer Epidemiology and Genetics, National Cancer Institute, Rockville, MD, USA*

NILANJAN CHATTERJEE *

*Departments of Biostatistics and Oncology, The Johns Hopkins University, Baltimore, MD,*

*USA*

nchatte2@jhu.edu

Summary

Prior to clinical applications, it is critical that risk prediction models are evaluated in independent studies that did not contribute to model development. While prospective cohort studies provide a natural setting for model validation, they often ascertain information on some risk factors (e.g., an expensive biomarker) in a nested sub-study of the original cohort, typically selected based on case-control status, and possibly some additional covariates. In this article, we propose an efficient approach for evaluating discriminatory ability of models using data from all individuals in a cohort study irrespective of whether they were sampled in the nested sub-study for measuring the complete set of risk factors. For evaluation of the Area Under the Curve (AUC) statistics, we estimate probabilities of risk-scores for cases being larger than those in controls conditional on partial risk-scores, the component of the risk-score that could be defined based on partial covariate information. The use of partial risk-scores, as opposed to actual multivariate risk-factor profiles, allows estimation of the underlying conditional expectations using subjects with complete covariate information in a non-parametric fashion even when numerous covariates are involved. We propose an influence function based approach for estimation of the variance of the resulting AUC statistics. We evaluate finite sample performance of the proposed method and compare it to an inverse probability weighted (IPW) estimator through extensive simulation studies. Finally, we illustrate an application of the proposed method for evaluating performance of a lung cancer risk prediction model using data from the Prostate, Lung, Colorectal and Ovarian Cancer (PLCO) trial.

*To whom correspondence should be addressed: nchatte2@jhu.edu

## 1. Introduction

The ultimate goal of a disease risk prediction model is to provide validated tools to patients and clinicians for making optimal healthcare decisions for disease prevention based on individual patient characteristics. For a risk prediction model to be clinically useful, it should be first evaluated in independent studies that did not contribute to the estimation of model parameters. Model evaluation involves two aspects: model calibration, i.e., evaluate whether the model is producing unbiased estimates of the disease risk for subjects with different risk factor profiles; and model discrimination, i.e., the ability of the model to discriminate between cases and controls (Gail and Pfeiffer, 2005; Chatterjee *and others*, 2016; Pfeiffer and Gail, 2017).

Prospective cohort studies, where disease-free individuals are enrolled and followed over time to observe disease incidence, provide an ideal setting for validation of risk models. However, cohort studies often collect data on risk factors using a two-phase design creating complex data structure. While all the cohort subjects contribute information on some epidemiologic risk factors (e.g., demographic factors like age, gender, race); information on some expensive covariates, such as a biomarker, are collected only on a nested sub-study of the original cohort. The sub-sample is typically stratified with respect to case/control status and certain other covariates that are observed for all the subjects in the cohort. The methods for analysis of data from such two phase studies for the purpose of fitting various regression models such as logistic regression or the Cox proportional hazard model have been well studied. (Prentice, 1986; Breslow and Cain, 1988; Langholz and Borgan, 1995; Reilley and Pepe, 1995; Scott and Wild, 1997; Langholz and Borgan, 1997; Breslow and Holubkov, 1997; Chatterjee *and others*, 2003). There has been, however, limited investigations on how the data from two-phase cohort studies can be efficiently used for model validation. In this article, we focus on evaluating discriminatory ability of models that can quantify risks of individuals through underlying risk-scores, i.e., weighted linear combinations of the risk factors. Such models may include commonly used parametric and semi-parametric models, such as logistic regression and Cox's proportional hazard model, where the effects of covariates are modeled parametrically and risk-scores could be defined by the underlying "linear predictor" of the model.

For models where disease risks are monotone functions of underlying risk-scores, the popular measure of Area Under the Curve (AUC) for evaluation of model discriminatory performance can be defined as the probability that the risk-score for a randomly selected subject with the disease is higher than that for a randomly selected subject without the disease. For studies that employ two-phase sampling, one can estimate AUC based only on the second phase study subjects that have complete risk factor information. In order to correct for the bias due to the complex sampling, each subject can be weighted by the inverse of probability of selection (Horvitz and Thompson, 1952). The probability of selection may be known or estimated from the data itself, possibly to increase efficiency of the analysis. Inverse probability weighted (IPW) estimators have been widely used for model evaluation in settings with complex sampling design (Cai and Zheng, 2011, 2012; Zheng *and others*, 2013; Zhou *and others*, 2013; Yao *and others*, 2015; Huang, 2016; Zheng *and others*, 2017). One can possibly conduct more efficient analysis of model evaluation using both phases of two-phase studies by imputation of missing risk-factor information for subjects not selected at the second phase (see e.g., Long *and others* (2011); Liu and Zhao (2012)). Such an approach, however, requires parametric model assumption for imputation distribution and may be particularly of concern in studies of model validation which are supposed to be empirical in nature.

In this article, we propose an alternative method of estimation of AUC that efficiently utilizes data from both phases of two-phase studies. Our method relies on the general principle of handling missing data in estimating equations using the conditional expectation operator. In this method, for the subjects who are not included in the second phase sample, we simply replace their contribution to the underlying estimating equation with the conditional expectation given their partial risk-scores, defined as observable component of the risk-scores of the underlying model. By conditioning on partial risk-scores, instead of observable risk-factor profiles, we achieve efficient dimension reduction without losing underlying risk information provided by the covariates. The dimension reduction allows estimation of conditional expectations, using data from subjects with complete risk-factor information, in a fairly non-parametric fashion even when numerous risk factors are evaluated in phase-I. We also derive an influence function based asymptotic variance estimate of our proposed estimator.

This paper is organized as follows: in Section 2, we describe our proposed statistical methodology and give an overview of the major results. In Section 3, we assess the finite sample performance of the proposed estimator and its standard error estimator and evaluate its efficiency relative to simple IPW estimator using extensive simulation studies. In Section 4, we present results from an application of the proposed method for evaluation of a lung cancer risk prediction model using data from a cohort study that evaluated important biomarkers for inflammation on a nested case-control study. In Section 5, we conclude the article with a discussion. The mathematical details of variance calculations, additional simulation results and additional details of real data analysis are provided in Supplementary Materials.

## 2. Methods

Suppose a study includes a total of $N_1$ cases and $N_0$ controls ($N = N_1 + N_0$). Let $X$ denote the design vector associated with a set of risk factors included in a risk prediction model for binary disease outcome and $\beta$ denote the corresponding vector of risk parameters. We assume that the model defines risks of individuals in terms of the risk-factors through an underlying linear predictor, or "risk-score", in the form $S = X\beta$. Here $X$ can include individual risk-factors and possibly their interaction terms as specified by the underlying model. We further assume that both the model and the underlying risk parameters ($\beta$) are pre-specified and have been derived based on analysis of prior studies and the goal of the current study is only to assess discriminatory power of the given model. We assume, some risk-factors (e.g., an expensive biomarker) may be measured only on a sub-sample of the subjects, which may have been selected based on case-control status and possibly some other covariate characteristics. The risk factors observed on all the subjects in the cohort are called phase-I risk factors and those that are observed only on the sub-sample are called phase-II risk factors. Accordingly we can partition $X$ as $X = (Z, W)$ and $\beta$ as $\beta = (\beta_Z, \beta_W)$, where Z denote the sub-vector of $X$ observed on all subjects in the study, $W$ denote the sub-vector of $X$ observed only on the subjects in the second phase sub-sample, $\beta_Z$ and $\beta_W$ are the risk parameters associated with $Z$ and $W$ respectively. Here, we note that, if model involves interaction terms among phase-I and phase-II risk factors, then $Z$, by definition, will not include elements for corresponding interaction terms because they are not "observable" at phase-I.

The Area Under the Curve (AUC) can be defined as $\delta_0 = P[S_1 > S_0]$, where $S_1$ and $S_0$ are the full risk-scores for the cases and controls respectively. Based on the observed and missing design vector profiles, we can decompose the risk-scores as $S = S_{obs} + S_{mis} = Z\beta_Z + W\beta_W$. For the cases and controls, we express these decompositions as: $S_1 = S_{1,obs} + S_{1,mis}$ and $S_0 = S_{0,obs} + S_{0,mis}$ respectively. Here, for cases(controls) included in the second phase sub-sample, $S_1(S_0)$ are observed; but for cases(controls) not included only $S_{1,obs}(S_{0,obs})$ are observed. Let $R_1(R_0)$ be the indicator of inclusion of a case(control) in the second phase sub-sample. We assume subjects may be sampled in the second-phase sample with sampling probability depending case-control status and possibly some covariates ($Z$) that are observed for subjects in the entire sample. Let $Z_1(Z_0)$ be the vector of these covariates for cases(controls) in the study. We further assume $\pi_1(Z_1) = P[R_1 = 1|Z_1]$ and $\pi_0(Z_0) = P[R_0 = 1|Z_0]$ defined as the sampling probabilities for the case and controls, respectively, to be known. Some of the covariates in $Z_1(Z_0)$ could be risk factors in the models and thus could be included in $S_{1,obs}(S_{0,obs})$.

When all the risk factors are observed on the entire cohort, i.e., there are no missing risk factors, the AUC can be estimated by the empirical proportion of case-control pairs for which the risk-score for the case is greater than that of the control. The underlying estimating equation can be written in the form: $\sum_{i=1}^{N_1}\sum_{j=1}^{N_0}(I(S_{1i} > S_{0j}) - \delta) = 0$. DeLong *and others* (1988) derive a variance formula for this estimate using general theory of U-statistics.

In the presence of two-phase sampling, the inverse probability weighted estimator of AUC can be obtained by solving the estimating equations: $\sum_{i=1}^{N_1}\sum_{j=1}^{N_0} \frac{R_{1i}}{\pi_1(Z_{1i})} \frac{R_{0j}}{\pi_0(Z_{0j})} \left(I(S_{1i} > S_{0j}) - \delta\right) = 0$, leading to the solution (Huang, 2016):

$$\hat{\delta}_{IPW} = \frac{\sum_{i=1}^{N_1}\sum_{j=1}^{N_0} \frac{R_{1i}}{\pi_1(Z_{1i})} \frac{R_{0j}}{\pi_0(Z_{0j})} I(S_{1i} > S_{0j})}{\sum_{i=1}^{N_1}\sum_{j=1}^{N_0} \frac{R_{1i}}{\pi_1(Z_{1i})} \frac{R_{0j}}{\pi_0(Z_{0j})}}$$

Suppose $N_1/N \to \lambda \in (0,1)$. Using empirical process theory (details in Supplementary Materials), the influence function representation of this estimator can be derived as:

$$(\hat{\delta}_{IPW} - \delta_0) = \frac{1}{N_1}\sum_{i=1}^{N_1} \frac{R_{1i}}{\pi_1(Z_{1i})}\left(p_0(S_{1i}) - \delta_0\right) + \frac{1}{N_0}\sum_{j=1}^{N_0} \frac{R_{0j}}{\pi_0(Z_{0j})}\left(p_1(S_{0j}) - \delta_0\right) + o_p(1/\sqrt{N}) \quad (2.1)$$

where $p_d(S_{1-d}) = E_{S_d}[I(S_1 > S_0)] = P_{S_d}[S_1 > S_0]$ for $d = 0, 1$. Consequently, the variance of $\hat{\delta}_{IPW}$ can be approximated as:

$$Var[\hat{\delta}_{IPW}] \approx \frac{1}{N_1} E\left[\left(\frac{R_1}{\pi_1(Z_1)}\left(p_0(S_1) - \delta_0\right)\right)^2\right] + \frac{1}{N_0} E\left[\left(\frac{R_0}{\pi_0(Z_0)}\left(p_1(S_0) - \delta_0\right)\right)^2\right] \quad (2.2)$$

A similar variance formula for the IPW estimator is given in Huang (2016). One may estimate the expectation in each term of (2.2) using the corresponding empirical weighted average using the information on second phase subjects and the sampling weights.

The inverse probability weighted approach can lead to efficiency loss in the AUC estimation by discarding the partial risk factor information from the subjects not included in the second phase sub-sample. To improve efficiency, we propose solving an alternative estimating equation of the form: $\sum_{i=1}^{N_1}\sum_{j=1}^{N_0}(\hat{\delta}_{ij} - \delta) = 0$, where $\hat{\delta}_{ij}$s are the estimators of $\delta_{ij}$s:

$$\begin{aligned}\delta_{ij} &= R_{1i}R_{0j}I(S_{1i} > S_{0j}) \\ &+ R_{1i}(1-R_{0j})E[I(S_{1i} > S_{0j})|S_{1i}, S_{0j,obs}] \\ &+ (1-R_{1i})R_{0j}E[I(S_{1i} > S_{0j})|S_{1i,obs}, S_{0j}] \\ &+ (1-R_{1i})(1-R_{0j})E[I(S_{1i} > S_{0j})|S_{1i,obs}, S_{0j,obs}]\end{aligned} \quad (2.3)$$

In (2.3), the decomposition of $\delta_{ij}$ into four components correspond to the inclusion status of the different case-control pairs to the second phase sub-sample. Each term corresponds to taking conditional expectation of the term, $I(S_1 > S_0)$, i.e., the indicator of whether the risk-score for a case is greater than that of the control, given the observed risk-scores for the case-control pairs. This approach based on conditioning on the observed risk-scores, and not the observed multivariate risk-factor profiles, allows us to estimate the conditional expectations in (2.3) in a fairly non-parametric fashion even when the number of observed risk-factors are relatively large. We propose to consider fine categories (e.g., deciles) of the observed risk-score and estimate the conditional expectations empirically based on the second phase sample accounting for sampling weights. In particular, the conditional expectations are estimated as follows:

$$\hat{E}[I(S_{1i} > S_{0j})|S_{1i}, S_{0j,obs}] = \frac{\sum_{l=1}^{N_0} \frac{R_{0l}}{\pi_0(Z_{0l})} I(S^c_{0l,obs} = S^c_{0j,obs}) I(S_{1i} > S_{0l})}{\sum_{l=1}^{N_0} \frac{R_{0l}}{\pi_0(Z_{0l})} I(S^c_{0l,obs} = S^c_{0j,obs})}$$

$$\hat{E}[I(S_{1i} > S_{0j})|S_{1i,obs}, S_{0j}] = \frac{\sum_{k=1}^{N_1} \frac{R_{1k}}{\pi_1(Z_{1k})} I(S^c_{1k,obs} = S^c_{1i,obs}) I(S_{1k} > S_{0j})}{\sum_{k=1}^{N_1} \frac{R_{1k}}{\pi_1(Z_{1k})} I(S^c_{1k,obs} = S^c_{1i,obs})}$$

$$\hat{E}[I(S_{1i} > S_{0j})|S_{1i,obs}, S_{0j,obs}] = \frac{\sum_{k=1}^{N_1}\sum_{l=1}^{N_0} \frac{R_{1k}R_{0l}}{\pi_1(Z_{1k})\pi_0(Z_{0l})} I(S^c_{1k,obs} = S^c_{1i,obs}, S^c_{0l,obs} = S^c_{0j,obs}) I(S_{1k} > S_{0l})}{\sum_{k=1}^{N_1}\sum_{l=1}^{N_0} \frac{R_{1k}R_{0l}}{\pi_1(Z_{1k})\pi_0(Z_{0l})} I(S^c_{1k,obs} = S^c_{1i,obs}, S^c_{0l,obs} = S^c_{0j,obs})}$$

where $S^c_{1,obs}$ and $S^c_{0,obs}$ are the categorical versions (e.g., deciles) of the observed risk-score for cases and controls in the cohort respectively. Plugging in these estimators we get $\hat{\delta}_{ij}$ and the two phase estimator of AUC can be written as:

$$\hat{\delta}_{TPS} = \frac{1}{N_1 N_0} \sum_{i=1}^{N_1} \sum_{j=1}^{N_0} \hat{\delta}_{ij}$$

The main result regarding the asymptotic theory of the estimator is summarized in the following theorem and a sketch of the proof is given in the Supplementary Materials.

THEOREM 2.1  Suppose $N_1/N \to \lambda \in (0,1)$. Let $p_d(S_{1-d})$ be as defined in (2.1), $p(S_{d,obs}) = E[p_d(S_{1-d})|S_{1-d,obs}]$, $p(S_d, S^c_{1-d,obs}) = E[I(S_1 > S_0)|S_d, S^c_{1-d,obs}]$, $p(S^c_{d,obs}, S^c_{1-d,obs}) = E[I(S_1 > S_0)|S^c_{d,obs}, S^c_{1-d,obs}]$, $\psi_d(S_d) = E_{Z_1, Z_0|S^c_{d,obs}}[(1-\pi_d(Z_d))(p(S_d, S^c_{1-d,obs}) - p(S^c_{d,obs}, S^c_{1-d,obs}))|S^c_{d,obs}]$ for $d = 0, 1$. Under mild regularity conditions the influence function representation of $\hat{\delta}_{TPS}$ can be written as:

$$(\hat{\delta}_{TPS} - \delta_0) \approx \frac{1}{N_1} \sum_{i=1}^{N_1} \left( R_{1i} p_0(S_{1i}) + (1 - R_{1i}) p(S_{1i,obs}) + \frac{R_{1i}}{\pi_{1i}(Z_{1i})} \psi_1(S_{1i}) - \delta_0 \right)$$
$$+ \frac{1}{N_0} \sum_{j=1}^{N_0} \left( R_{0j} p_1(S_{0j}) + (1 - R_{0j}) p(S_{0j,obs}) + \frac{R_{0j}}{\pi_{0j}(Z_{0j})} \psi_0(S_{0j}) - \delta_0 \right) + o_p(1/\sqrt{N})$$
(2.4)

From (2.4), the asymptotic variance of $\hat{\delta}_{TPS}$ can be expressed as:

$$Var[\hat{\delta}_{TPS}] \approx \frac{1}{N_1} E\left[ \left( R_1 p_0(S_1) + (1 - R_1) p(S_{1,obs}) + \frac{R_1}{\pi_1(Z_1)} \psi_1(S_1) - \delta_0 \right)^2 \right]$$
$$+ \frac{1}{N_0} E\left[ \left( R_0 p_1(S_0) + (1 - R_0) p(S_{0,obs}) + \frac{R_0}{\pi_0(Z_0)} \psi_0(S_0) - \delta_0 \right)^2 \right] \quad (2.5)$$

In (2.5), the two terms correspond to contributions from the cases and controls respectively. Each subject included in the second phase sample contributes twice to the variability: the first one is due to their direct contribution to the estimation of the AUC and the second one comes from their contribution to estimation of the conditional expectations described earlier.

Each expectation term in (2.5) can be estimated using the corresponding empirical average. Estimates of $p_0(S_1), p_1(S_0), p(S_{1,obs}), p(S_{0,obs}), \psi_1(S_1), \psi_0(S_0)$ are obtained from the subjects in the second phase sample by the corresponding weighted averages over case-control pairs within stratum defined by the observed risk-score categories using the sampling weights.

## 3. SIMULATION STUDY

We conducted extensive simulation studies to assess the finite sample performance of the proposed estimator in terms of bias, variance and coverage of 95% Wald based confidence intervals and also relative efficiency gain compared to the IPW estimator.

3.1  *Simulation Design*

We simulate data for a cohort study of N=50,000 individuals based on a model involving a total of 8 independent risk-factors. We assume four of the risk-factors ($X_1$, $X_2$, $X_3$ and $X_4$) are continuously distributed as standard normal variates, and the other four are binary ($X_5$, $X_6$, $X_7$ and $X_8$), distributed as Bernoulli random variables each with success probability $p = 0.5$. Let $X = (X_1, X_2, X_3, X_4, X_5, X_6, X_7, X_8)$. Let $\beta$ be the vector of log-relative risks associated with these covariates. For each individual we generate an age of disease onset $T$ from a Cox model of the form $\lambda(t|X = x) = \lambda_0(t)exp(x\beta)$, where $\lambda_0(t) = \lambda\gamma t^{\gamma-1}$ is the hazard function corresponding to $Weibull(\lambda, \gamma)$ distribution. The log-relative risk parameters of the model were chosen to be $\beta = (\log(1.1), \log(1.1), \log(1.1), \log(1.1), \log(1.2), \log(1.2), -\log(1.2), -\log(1.2))^T$, which correspond to moderate level of associations for each of the individual risk-factors with the disease. Moreover, the parameters associated with the baseline hazard were chosen such that the probability of developing the disease by age 50 and 70 years are approximately 5% and 12% respectively. We generate age of entry for a subject from a discrete uniform distribution in the range 50 years to 70 years. We also generate a length of observed follow-up from a discrete uniform distribution in the range 19 years to 21 years. The data generating mechanism assumes that each subject is disease free at the time of entry into the cohort.

After we simulate data on full cohort, we select a nested case-control sample using alternative designs and assume that data on two of the risk-factors, $X_7$ and $X_8$, are only available on the subsample. Thus, we assume $X = (Z, W)$, where $Z = (X_1, X_2, X_3, X_4, X_5, X_6)$ and $W = (X_7, X_8)$. Accordingly, $\beta$ can be partitioned as $\beta = (\beta_Z, \beta_W)$ and we define $S_{obs} = Z\beta_Z$, $S_{mis} = W\beta_W$ and $S = S_{obs} + S_{mis}$. We further define $f$ to be the variance of $S_{obs}$, the partial risk-score, as a ratio to that of $S$, the total risk-score, and use it as an index to quantify the relative importance of the phase-I risk-factors compared to the entire set of risk-factors. For our chosen $\beta = (\log(1.1), \log(1.1), \log(1.1), \log(1.1), \log(1.2), \log(1.2), -\log(1.2), -\log(1.2))^T$, we have $f \approx$ 0.75. We vary value of $f$ in the set 0.05,0.1,0.2,0.4,0.5,0.75 by varying the underlying values of the log-relative-risk parameters of the risk factors and the parameters of the baseline hazard in the model.

We explore two sampling strategies to simulate the subjects in the second phase nested case-control sample: simple case-control sampling and stratified case-control sampling.

*Simple case-control sampling:* Here we create a nested case-control study by selecting random samples of cases and controls from the cohort. We vary the fraction of cases to be sampled in the second phase study in the range $\eta = 1, 0.75, 0.5, 0.25, 0.1$. In each setting, we sample roughly equal number of controls. The sampling probability of the controls were empirically estimated as the number of controls sampled as a ratio to the total number in the full cohort.

*Stratified case-control sampling:* Under this setting, we consider a sampling scheme where cases and controls were sampled in the second-phase by matching with respect to the partial risk-scores that could be defined based on the risk-factors observed in the entire cohort. We first stratify the cohort into ten categories based on deciles of the observed risk-scores among cases and then draw roughly equal number of cases and controls from each category. As before, we vary $\eta$, the proportion of cases to be sampled at the second phase, over a range. Under this design, we also determine the sampling probabilities of the controls empirically from our simulation studies.

For each simulation setting, we repeatedly generate the cohort study with complete risk factor information to obtain 1000 simulated cohort datasets. We compute the AUC from each dataset as the empirical proportion of case-control pairs for which the risk-score for the case is greater than that of the control. We compute an average of these 1000 AUC values and consider it to be the true AUC of the underlying model.

### 3.2 *Simulation Results*

Table 1 shows the simulation results evaluating the performance of the proposed and the IPW estimators under alternative sampling schemes with sampling fraction for cases $1, 0.5, 0.25$ and $f = 0.75$. In all the scenarios, both the estimators have very small bias and the confidence intervals (constructed using influence function based variance estimates) achieve the nominal 95% level. The percent bias in the standard error estimate is also very small. In all the scenarios, the proposed estimator is much more efficient compared to the IPW estimator. When the second phase sample includes all the cases and a random sample of the controls, the proposed approach leads to approximately 50% efficiency gain. Compared to simple case-control sampling, the straified case-control sampling of the second phase subjects leads to modest efficiency loss. Similar simulation tables for $f = 0.5$ and $f = 0.2$ are shown in the Supplementary Materials.

Figure 1 shows the relative efficiency of the proposed estimator compared to the IPW estimator as a function of the fraction of cases sampled ($\eta$) under simple case-control sampling of the second phase subjects. For fixed $f$, the relative efficiency increases as the fraction of cases sampled at phase II decreases because the IPW estimator fails to incorporate information from the increasing number of unselected subjects. Moreover, for fixed fraction of cases sampled, the relative efficiency increases with increase in $f$ as the observed risk-score explains a larger proportion of the variability of the full risk-score and the proposed estimator gains efficiency by using the observed risk-score from all the subjects in the cohort.

## 4. Data Analysis

We illustrate an application of the proposed method to evaluate the discriminatory performance of a lung cancer risk prediction model that has a potential clinical application for selecting subjects for CT-screening. The CT scan procedure, which uses low-dose radiation from x-ray machines to scan the body in a helical path and produce detailed image of regions inside the body, has been demonstrated as an effective method for reducing lung cancer mortality compared to chest radiography (National Lung Screening Trial Research Team *and others*, 2011; Gould, 2014). As a consequence, the US Preventive Services Task Force (USPSTF) recommended annual CT screening for lung cancer in certain risk factor based subgroups of individuals (Moyer, 2014; de Koning *and others*, 2014; McMahon *and others*, 2014). However, screening is likely to be more beneficial if the subjects at higher risk of lung cancer can be identified based on individual risk predictions (i.e., risk based selection) (Bach and Gould, 2012; Bach, 2014; Gould, 2014; Tammemägi, 2015). This can be implemented by developing and validating a comprehensive lung cancer risk prediction model that includes the major risk factors.

The original model included a host of epidemiologic risk factors: gender, race, education, BMI, smoking pack-years coded as a categorical variable, number of years since stopped smoking cigarettes, number of years smoked, binary indicator of $> 1$ pack/day, presence/absence of emphysema, lung cancer family history (Katki *and others*, 2016). The model was developed using data from the control arm of the Prostate, Lung, Colorectal, and Ovarian Cancer Screening Trial (PLCO: 1993-2009), which included approximately 39,000 ever smokers in the age range 55-74 years with epidemiologic risk factor information. The variable coding and relative risk information for these risk factors are provided in Table 3 of Supplementary Materials.

We wanted to evaluate the potential added value of certain inflammation biomarkers into the model using data that has been collected within the screening arm of the PLCO study. Information on four biomarkers, C-Reactive Protein [CRP], serum amyloid A [SAA], soluble tumor necrosis factor receptor 2 [sTNFRII] and monokine induced by gamma interferon [CXCL9/MIG], were collected in two nested case-control studies within the screening arm of the PLCO study: the discovery study and the replication study (Shiels *and others*, 2013, 2015). The discovery study included 960 ever smokers aged 55-74 years (500 cases and 460 controls) and the replication study included 929 ever smokers aged 55-74 years (468 cases and 461 controls). Among these biomarkers,

CRP has been shown to be consistently associated with lung cancer (Chaturvedi and others, 2010; Pine and others, 2011; Leuzzi and others, 2016), while the other three were identified as most promising ones among a larger set of biomarkers studied in the discovery and replication samples. We obtained the relative risk parameters of the four biomarkers, after adjustment of the other epidemiologic factors, using the discovery sample (see Table 3 of Supplementary Materials for coding of these variable and estimates of associated risk parameters). These estimates together with those obtained for the traditional epidemiologic factors from the control arm of the PLCO study provided a combined model for predicting lung cancer risk based on the epidemiologic risk factors and biomarkers.

In the current analysis, we evaluated discriminatory accuracy of the combined model using the screening arm of the PLCO study and the replication biomarker study. Since neither the epidemiologic risk-factors from the screening cohort nor the biomarker information from the replication case-control study was utilized for model building, the resulting two-phase study can be assumed to be independent of model development.

We divide the PLCO screening arm participants into four age categories: $\leqslant 59$ years, 60-64 years, 65-69 years, $\geqslant 70$ years and evaluate the discriminatory performance in these four sub-cohorts defined by those categories. Table 2 shows the AUC estimates and the associated standard errors in the four sub-cohorts based on three combinations of risk factors: (i) epidemiologic risk factors only, (ii) epidemiologic risk factors and C-reactive protein and (iii) epidemiologic risk factors and all four inflammation biomarkers. For computing the proposed AUC estimator, we stratify the partial risk-score observed in phase-I into six categories based on sextiles. For the epidemiologic risk factor only model, we evaluated standard AUC estimator based on all smokers in the entire PLCO screening arm (Phase I sample). From results reported in Table 2 we observe that AUC estimates for each type of model substantially decreased from younger to the older age groups. For risk factor combinations (ii) and (iii), comparison of the proposed and IPW estimators shows that while point estimates were similar, the precision of the proposed estimator was much higher across all the settings. We observe that the inclusion of inflammation biomarkers leads to limited improvement of AUC irrespective of the age groups.

## 5. Discussion

In this article, we propose an efficient and robust estimator of AUC statistics in the setting of two-phase study designs. Our method combines complete risk factor information from the second phase sample and partial risk factor information from subjects not included in the second phase. We have derived an explicit formula for asymptotic variance of the proposed estimator. The results from the simulation studies and the real data analysis demonstrate the efficiency gain that our method achieves compared to the IPW approach.

The estimation of conditional probabilities requires adjustment for potential non-random sampling of the second phase subjects. We perform this step using inverse probability weighting assuming the selection probabilities (i.e., sampling weights) are known. In real data settings, it may be complicated to obtain these selection probabilities and they may need to be obtained through post-hoc estimation under parametric assumptions on the selection mechanism. In our real data application, we use sampling weights derived from a logistic regression model of inclusion on case/control status and other covariates (see Supplementary Materials). Our method may be subject to bias if the true selection mechanism does not belong to the class of parametric models considered for post-hoc estimation. Even when the selection model is correct, the method needs to appropriately account for the uncertainty of the post-hoc estimation of selection probabilities.

The setting we consider has a close connection with a recent study conducted by Zheng *and others* (2017). In a time-to event setting, they propose a weighted semiparametric likelihood approach to estimate the relative risks of the risk factors and the model evaluation statistics (e.g., time dependent AUC) using information from the second phase subjects. The method incorpo-rates information from Phase I subjects through estimation of selection probabilities conditional on observed risk-factor information. The non-parametric estimation approach they consider, however, could break down due to curse of dimensionality when numerous covariates are evaluated at the Phase I sample itself. We consider a simpler setting where we use only the disease status information of each subject in the cohort ignoring the information on follow-up time. In the future, we plan to extend our partial risk-score based approach to evaluate the AUC and possibly other model evaluation statistics in a time-to-event setting taking into account censoring and follow-up information.

In summary, the strengths of our approach are its practical usefulness and ease of implementation with numerous risk factors. The efficient dimension reduction achieved through partial risk-scores allows us to estimate model evaluation statistics in a non-parametric manner, which is particularly useful in model validation studies that are supposed to be empirical in nature.

## 6. Supplementary Material

The reader is referred to the Supplementary Materials for technical appendices, additional simulations and additional details of real data application.

## Acknowledgments

This work was supported by the Patient-Centered Outcomes Research Institute (PCORI) Award (ME-1602-34530). The statements and opinions in this article are solely the responsibility of the authors and do not necessarily represent the views of the Patient-Centered Outcomes Research Institute (PCORI), its Board of Governors, or Methodology Committee. *Conflict of Interest*: None declared.

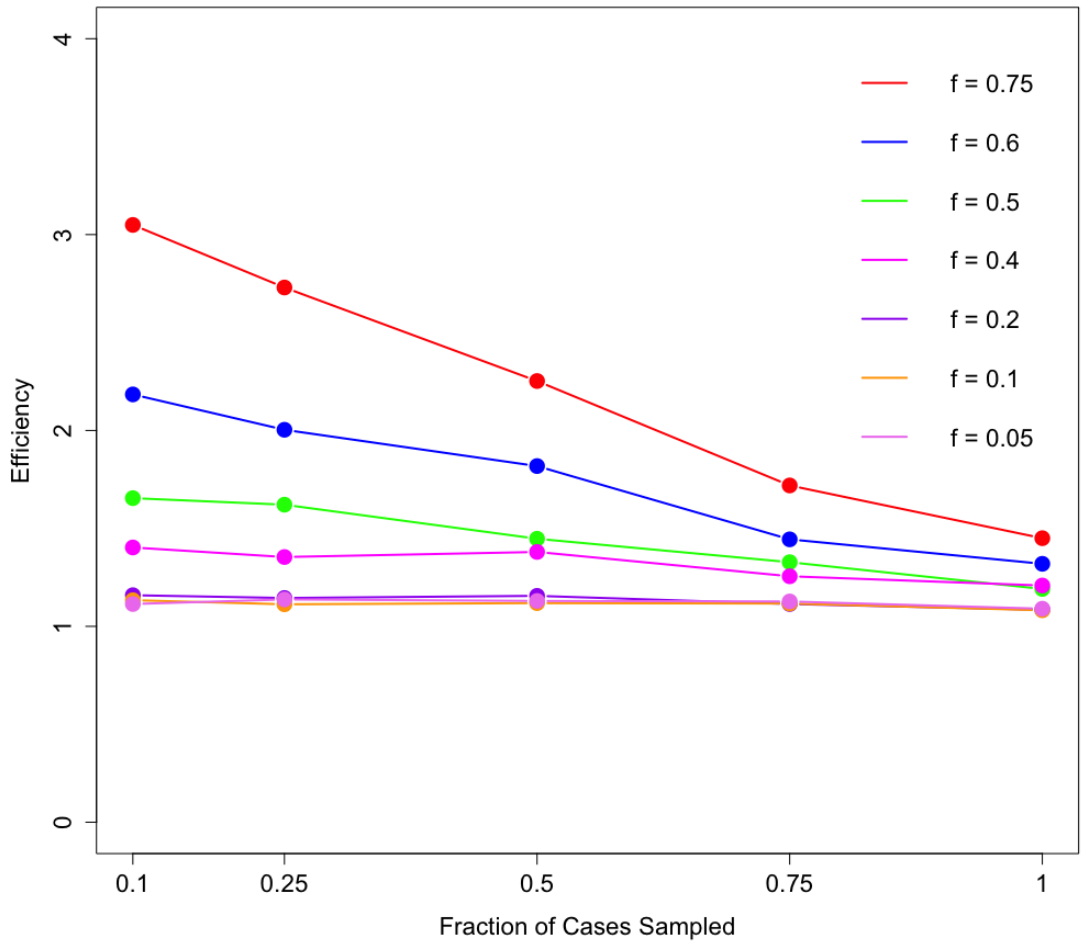

Fig. 1. Relative efficiency of our proposed estimator compared to the IPW estimator as a function of the fraction of cases sampled ($\eta$) under random case-control sampling of the second phase subjects. Different lines correspond to different values of $f$ characterizing different proportions of variability of the full risk-score explained by the observed risk-score.

Table 1. *Simulation studies evaluating performance of the proposed and the IPW estimators of AUC estimation under two-phase studies. Results are shown under alternative sampling schemes with varying sampling probabilities and a fixed value of $f = 0.75$, proportion of variability of the full risk score explained by the observed risk score. The true value of AUC of the underlying model is 0.577. Relative efficiency (RE) of the AUC estimates are reported in reference to standard estimate of AUC obtained using information on all risk factors for all subjects in the cohort. Percent bias in the estimation of standard errors of AUC using the influence function based variance estimate are also reported.*

| Sampling scheme | Bias | | RE (compared to full cohort) | | Percent bias in SE estimate | | Coverage | |
|---|---|---|---|---|---|---|---|---|
| | TPS | IPW | TPS | IPW | TPS | IPW | TPS | IPW |
| Simple case-control sampling | | | | | | | | |
| Sampling fraction of cases 1 | | | | | | | | |
| Case-control ratio 1:1 | $9.56 \times 10^{-5}$ | 0.0002 | 0.89 | 0.58 | -3.92 | -3.17 | 0.96 | 0.94 |
| Sampling fraction of cases 0.5 | | | | | | | | |
| Case-control ratio 1:1 | 0.0002 | $-4.5 \times 10^{-5}$ | 0.6 | 0.28 | -1.61 | -1.1 | 0.95 | 0.96 |
| Sampling fraction of cases 0.25 | | | | | | | | |
| Case-control ratio 1:1 | 0.0002 | 0.0004 | 0.41 | 0.14 | -5.33 | -0.78 | 0.96 | 0.95 |
| Stratified case-control sampling | | | | | | | | |
| Sampling fraction of cases 1 | | | | | | | | |
| Case-control ratio 1:1 | $7.08 \times 10^{-5}$ | $-6.39 \times 10^{-5}$ | 0.73 | 0.5 | 5.36 | 2.94 | 0.94 | 0.94 |
| Sampling fraction of cases 0.5 | | | | | | | | |
| Case-control ratio 1:1 | $5.74 \times 10^{-5}$ | 0.0001 | 0.6 | 0.28 | -1.61 | -3.33 | 0.96 | 0.96 |
| Sampling fraction of cases 0.25 | | | | | | | | |
| Case-control ratio 1:1 | -0.0002 | 0.0001 | 0.4 | 0.13 | -3.95 | 0 | 0.96 | 0.95 |

Table 2. *Age category specific AUC estimates and standard errors for three alternative risk models: (i) epidemiologic risk factors only, (ii) epidemiologic risk factors and C-reactive protein, (iii) epidemiologic risk factors and four inflammation markers. Phase I subjects are the ever-smokers aged 55-74 years in the screening arm of the PLCO Trial; Phase II subjects are those ever-smokers aged 55-74 years included in the replication study (nested case-control study within PLCO screening arm). For epidemiologic risk factor only model, the standard AUC estimator based on Phase I sample is reported.*

| | | | AUC estimate (standard deviation) | | | | |
|---|---|---|---|---|---|---|---|
| | Number of subjects (Number of cases) | | Epidemiologic risk factors only | Epidemiologic risk factors + CRP | | Epidemiologic risk factors + 4 inflammation markers | |
| Age (in years) | Phase I | Phase II | | IPW | TPS | IPW | TPS |
| ⩽ 59 | 13,443 (321) | 208 (103) | 0.799 (0.011) | 0.777 (0.047) | 0.803 (0.028) | 0.78 (0.046) | 0.804 (0.027) |
| 60-64 | 12,054 (459) | 277 (140) | 0.761 (0.011) | 0.766 (0.031) | 0.752 (0.023) | 0.762 (0.031) | 0.75 (0.023) |
| 65-69 | 8669 (489) | 277 (140) | 0.777 (0.01) | 0.802 (0.034) | 0.801 (0.023) | 0.8 (0.035) | 0.8 (0.024) |
| ⩾ 70 | 4632 (277) | 167 (85) | 0.709 (0.015) | 0.707 (0.05) | 0.715 (0.026) | 0.719 (0.049) | 0.725 (0.025) |
| Total | 38,798 (1546) | 929 (468) | | | | | |

# Evaluating discriminatory accuracy of models using partial risk-scores in two-phase studies: Supplementary Materials

PARICHOY PAL CHOUDHURY

*Department of Biostatistics, The Johns Hopkins University, Baltimore, MD, USA*

parichoy@jhu.edu

ANIL K. CHATURVEDI

*Division of Cancer Epidemiology and Genetics, National Cancer Institute, Rockville, MD, USA*

NILANJAN CHATTERJEE *

*Departments of Biostatistics and Oncology, The Johns Hopkins University, Baltimore, MD,*

*USA*

nchatte2@jhu.edu

1. VARIANCE OF THE TWO-PHASE ESTIMATOR

In this section we give a sketch of proof for Theorem 2.1 of the manuscript. We derive the influence function representation of $\hat{\delta}_{TPS}$ using the following steps:

**Step 1:** Note that:

$$(\hat{\delta}_{TPS} - \delta_0) = \frac{1}{N_1 N_0} \sum_{i=1}^{N_1} \sum_{j=1}^{N_0} (\delta_{ij} - \delta_0) + \frac{1}{N_1 N_0} \sum_{i=1}^{N_1} \sum_{j=1}^{N_0} (\hat{\delta}_{ij} - \delta_{ij}) \qquad (1.1)$$

*To whom correspondence should be addressed: nchatte2@jhu.edu

**Step 2:** The first term in equation 1.1 can be expressed as:

$$\frac{1}{N_1 N_0} \sum_{i=1}^{N_1} \sum_{j=1}^{N_0} (\delta_{ij} - \delta_0)$$

$$= \frac{1}{N_1 N_0} \sum_{i=1}^{N_1} \sum_{j=1}^{N_0} R_{1i} R_{0j} (I(S_{1i} > S_{0j}) - \delta_0)$$

$$+ \frac{1}{N_1 N_0} \sum_{i=1}^{N_1} \sum_{j=1}^{N_0} R_{1i}(1 - R_{0j})(E[I(S_{1i} > S_{0j})|S_{0i}, S_{0j,obs}] - \delta_0)$$

$$+ \frac{1}{N_1 N_0} \sum_{i=1}^{N_1} \sum_{j=1}^{N_0} (1 - R_{1i}) R_{0j} (E[I(S_{1i} > S_{0j})|S_{1i,obs}, S_{0j}] - \delta_0)$$

$$+ \frac{1}{N_1 N_0} \sum_{i=1}^{N_1} \sum_{j=1}^{N_0} (1 - R_{1i})(1 - R_{0j})(E[I(S_{1i} > S_{0j})|S_{1i,obs}, S_{0j,obs}] - \delta_0) \quad (1.2)$$

The first term in equation 1.2 can be expressed as:

$$\frac{1}{N_1 N_0} \sum_{i=1}^{N_1} \sum_{j=1}^{N_0} R_{1i} R_{0j}(I(S_{1i} > S_{0j}) - \delta_0)$$

$$= \int_{(R_1, S_1)} \int_{(R_0, S_0)} R_1 R_0 (I(S_1 > S_0) - \delta_0) dF_{(R_1, S_1)}^{(N_1)} dF_{(R_0, S_0)}^{(N_0)}$$

$$\approx \int_{(R_1, S_1)} \int_{(R_0, S_0)} R_1 R_0 (I(S_1 > S_0) - \delta_0) dF_{(R_0, S_0)}^{(N_0)} dF_{(R_1, S_1)}^{(N_1)}$$

$$+ \int_{(R_0, S_0)} \int_{(R_1, S_1)} R_1 R_0 (I(S_1 > S_0) - \delta_0) dF_{(R_1, S_1)} dF_{(R_0, S_0)}^{(N_0)} \quad \text{by von Mises expansion}$$

$$= \frac{1}{N_1} \sum_{i=1}^{N_1} R_{1i} E_{Z_0}[\pi_0(Z_0) E_{S_{0,mis}|S_{0,obs}}[(I(S_{1i} > S_0) - \delta_0)|S_{0,obs}]]$$

$$+ \frac{1}{N_0} \sum_{j=1}^{N_0} R_{0j} E_{Z_1}[\pi_1(Z_1) E_{S_{1,mis}|S_{1,obs}}[(I(S_1 > S_{0j}) - \delta_0)|S_{1,obs}]]$$

The second term of equation 1.2 can be expressed as:

$$\frac{1}{N_1 N_0} \sum_{i=1}^{N_1} \sum_{j=1}^{N_0} R_{1i}(1 - R_{0j})(E[I(S_{1i} > S_{0j})|S_{1i}, S_{0j,obs}] - \delta_0)$$

$$= \int_{(R_1, S_1)} \int_{(R_0, S_{0,obs})} R_1(1 - R_0)(E[I(S_1 > S_0)|S_1, S_{0,obs}] - \delta_0) dF_{(R_1, S_1)}^{(N_1)} dF_{(R_0, S_{0,obs})}^{(N_0)}$$

$$\approx \int_{(R_1, S_1)} \int_{(R_0, S_{0,obs})} R_1(1 - R_0)(E[I(S_1 > S_0)|S_1, S_{0,obs}] - \delta_0) dF_{(R_0, S_{0,obs})} dF_{(R_1, S_1)}^{(N_1)}$$

$$+ \int_{(R_0, S_{0,obs})} \int_{(R_1, S_1)} R_1(1 - R_0)(E[I(S_1 > S_0)|S_1, S_{0,obs}] - \delta_0) dF_{(R_1, S_1)} dF_{(R_0, S_{0,obs})}^{(N_0)}$$

$$= \frac{1}{N_1} \sum_{i=1}^{N_1} R_{1i} E_{Z_0}[(1 - \pi_0(Z_0)) E_{S_{0,mis}|S_{0,obs}}[(I(S_{1i} > S_0) - \delta_0)|S_{0,obs}]]$$

$$+ \frac{1}{N_0} \sum_{j=1}^{N_0} (1 - R_{0j}) E_{Z_1}[\pi_1(Z_1) E_{S_{1,mis}|S_{1,obs}}[E_{S_{0,mis}|S_{0,obs}}[(I(S_1 > S_0) - \delta_0)|S_{0j,obs}]|S_{1,obs}]]$$

The third term of equation 1.2 can be expressed as:

$$\frac{1}{N_1 N_0} \sum_{i=1}^{N_1} \sum_{j=1}^{N_0} (1 - R_{1i}) R_{0j} (E[I(S_{1i} > S_{0j})|S_{1i,obs}, S_{0j}] - \delta_0)$$

$$= \int_{(R_1, S_{1,obs})} \int_{(R_0, S_0)} (1 - R_1) R_0 (E[I(S_1 > S_0)|S_{1,obs}, S_0] - \delta_0) dF_{(R_1, S_{1,obs})}^{(N_1)} dF_{(R_0, S_0)}^{(N_0)}$$

$$\approx \int_{(R_1, S_{1,obs})} \int_{(R_0, S_0)} (1 - R_1) R_0 (E[I(S_1 > S_0)|S_{1,obs}, S_0] - \delta_0) dF_{(R_0, S_0)} dF_{(R_1, S_{1,obs})}^{(N_1)}$$

$$+ \int_{(R_0, S_0)} \int_{(R_1, S_{1,obs})} (1 - R_1) R_0 (E[I(S_1 > S_0)|S_{1,obs}, S_0] - \delta_0) dF_{(R_1, S_{1,obs})} dF_{(R_0, S_0)}^{(N_0)}$$

$$= \frac{1}{N_1} \sum_{i=1}^{N_1} (1 - R_{1i}) E_{Z_0} [\pi_0(Z_0) E_{S_{0,mis}|S_{0,obs}} [E_{S_{1,mis}|S_{1,obs}} [(I(S_{1i} > S_0) - \delta_0)|S_{1i,obs}]|S_{0,obs}]]$$

$$+ \frac{1}{N_0} \sum_{j=1}^{N_0} R_{0j} E_{Z_1} [(1 - \pi_1(Z_1)) E_{S_{1,mis}|S_{1,obs}} [(I(S_1 > S_{0j}) - \delta_0)|S_{1,obs}]]$$

The fourth term of equation 1.2 can be expressed as:

$$\frac{1}{N_1 N_0} \sum_{i=1}^{N_1} \sum_{j=1}^{N_0} (1 - R_{1i})(1 - R_{0j})(E[I(S_{1i} > S_{0j})|S_{1i,obs}, S_{0j,obs}] - \delta_0)$$

$$= \int_{(R_1, S_{1,obs})} \int_{(R_0, S_{0,obs})} (1 - R_1)(1 - R_0)(E[I(S_1 > S_0)|S_{1,obs}, S_{0,obs}] - \delta_0) dF_{(R_1, S_{1,obs})}^{(N_1)} dF_{(R_0, S_{0,obs})}^{(N_0)}$$

$$= \int_{(R_1, S_{1,obs})} \int_{(R_0, S_{0,obs})} (1 - R_1)(1 - R_0)(E[I(S_1 > S_0)|S_{1,obs}, S_{0,obs}] - \delta_0) dF_{(R_0, S_{0,obs})} dF_{(R_1, S_{1,obs})}^{(N_1)}$$

$$+ \int_{(R_0, S_{0,obs})} \int_{(R_1, S_{1,obs})} (1 - R_1)(1 - R_0)(E[I(S_1 > S_0)|S_{1,obs}, S_{0,obs}] - \delta_0) dF_{(R_1, S_{1,obs})} dF_{(R_0, S_{0,obs})}^{(N_0)}$$

$$= \frac{1}{N_1} \sum_{i=1}^{N_1} (1 - R_{1i}) E_{Z_0} [(1 - \pi_0(Z_0)) E_{S_{0,mis}|S_{0,obs}} [E_{S_{1,mis}|S_{1,obs}} [(I(S_{1i} > S_0) - \delta_0)|S_{1i,obs}]|S_{0,obs}]]$$

$$+ \frac{1}{N_0} \sum_{j=1}^{N_0} (1 - R_{0j}) E_{Z_1} [(1 - \pi_1(Z_1)) E_{S_{1,mis}|S_{1,obs}} [(E_{S_{0,mis}|S_{0,obs}} [I(S_1 > S_{0j}) - \delta_0)|S_{0j,obs}]|S_{1,obs}]]$$

Adding up all the terms leads to:

$$\frac{1}{N_1 N_0} \sum_{i=1}^{N_1} \sum_{j=1}^{N_0} (\delta_{ij} - \delta_0)$$

$$= \frac{1}{N_1} \sum_{i=1}^{N_1} R_{1i}(p_0(S_{1i}) - \delta_0) + \frac{1}{N_0} \sum_{j=1}^{N_0} R_{0j}(p_1(S_{0j}) - \delta_0)$$

$$+ \frac{1}{N_1} \sum_{i=1}^{N_1} (1 - R_{1i})(p(S_{1i,obs}) - \delta_0) + \frac{1}{N_0} \sum_{i=1}^{N_0} (1 - R_{0j})(p(S_{0j,obs}) - \delta_0)$$

where $p_d(S_{1-d}) = E_{S_d}[I(S_1 > S_0)] = P_{S_d}[S_1 > S_0]$, $p(S_{d,obs}) = E[p_d(S_{1-d})|S_{d,obs}]$ for $d = 0, 1$.

We assume: $p(S_{d,obs}) \approx p(S^c_{d,obs})$ for $d = 0, 1$.

**Step 3:** The second term in equation 1.1 can be expressed as:

$$\frac{1}{N_1 N_0} \sum_{i=1}^{N_1} \sum_{j=1}^{N_0} (\hat{\delta}_{ij} - \delta_{ij})$$

$$= \frac{1}{N_1 N_0} \sum_{i=1}^{N_1} \sum_{j=1}^{N_0} R_{1i}(1 - R_{0j}) \left( \frac{\sum_{l=1}^{N_0} \frac{R_{0l}}{\pi_0(Z_{0l})} I(S^c_{0l,obs} = S^c_{0j,obs})\{I(S_{1i} > S_{0l}) - p(S_{1i}, S_{0j,obs})\}}{\sum_{l=1}^{N_0} \frac{R_{0l}}{\pi_0(Z_{0l})} I(S^c_{0l,obs} = S^c_{0j,obs})} \right)$$

$$+ \frac{1}{N_1 N_0} \sum_{i=1}^{N_1} \sum_{j=1}^{N_0} (1 - R_{1i}) R_{0j} \left( \frac{\sum_{k=1}^{N_1} \frac{R_{1k}}{\pi_1(Z_{1k})} I(S^c_{1k,obs} = S^c_{1i,obs})\{I(S_{1k} > S_{0j}) - p(S_{1i,obs}, S_{0j})\}}{\sum_{k=1}^{N_1} \frac{R_{1k}}{\pi_1(Z_{1k})} I(S^c_{1k,obs} = S^c_{1i,obs})} \right)$$

$$+ \frac{1}{N_1 N_0} \sum_{i=1}^{N_1} \sum_{j=1}^{N_0} (1 - R_{1i})(1 - R_{0j})$$

$$\times \left( \frac{\sum_{k=1}^{N_1} \sum_{l=1}^{N_0} \frac{R_{1k} R_{0l}}{\pi_1(Z_{1k})\pi_0(Z_{0l})} I(S^c_{1k,obs} = S^c_{1i,obs}, S^c_{0l,obs} = S^c_{0j,obs})\{I(S_{1k} > S_{0l}) - p(S_{1i,obs}, S_{0j,obs})\}}{\sum_{k=1}^{N_1} \sum_{l=1}^{N_0} \frac{R_{1k} R_{0l}}{\pi_1(Z_{1k})\pi_0(Z_{0l})} I(S^c_{1k,obs} = S^c_{1i,obs}, S^c_{0l,obs} = S^c_{0j,obs})} \right)$$

(1.3)

Define: $p(S_1, S_{0,obs}) = E[I(S_1 > S_0)|S_1, S_{0,obs}]$, $p(S_{1,obs}, S_0) = E[I(S_1 > S_0)|S_{1,obs}, S_0]$, $p(S_{1,obs}, S_{0,obs}) = E[I(S_1 > S_0)|S_{1,obs}, S_{0,obs}]$. We assume the following: $p(S_1, S_{0,obs}) \approx p(S_1, S^c_{0,obsc})$, $p(S_{1,obs}, S_0) \approx p(S^c_{1,obs}, S_0)$ and $p(S_{1,obs}, S_{0,obs}) \approx p(S^c_{1,obs}, S^c_{0,obs})$. Let $q_d(s^c_{d,obs}) = E[I(S^c_{d,obs} = s^c_{d,obs})] = P[S^c_{d,obs} = s^c_{d,obs}]$ for $d = 0, 1$. Note that $E[p(S_d, S_{1-d,obs})|S_{d,obs}, S_{1-d,obs}] = p(S_{d,obs}, S_{1-d,obs})$ for $d = 0, 1$.

The ratio term within first bracket in the first term in equation 1.3 can be expressed as:

$$\frac{\sum_{l=1}^{N_0} \frac{R_{0l}}{\pi_0(Z_{0l})} I(S^c_{0l,obs} = S^c_{0j,obs})\{I(S_{1i} > S_{0l}) - p(S_{1i}, S_{0j,obs})\}}{\sum_{l=1}^{N_0} \frac{R_{0l}}{\pi_0(Z_{0l})} I(S^c_{0l,obs} = S^c_{0j,obs})}$$

$$\approx \frac{1}{q_0(S^c_{0j,obs})} \frac{1}{N_0} \sum_{l=1}^{N_0} \frac{R_{0l}}{\pi_0(Z_{0l})} I(S^c_{0l,obs} = S^c_{0j,obs})\{I(S_{1i} > S_{0l}) - p(S_{1i}, S_{0j,obs})\} \quad (1.4)$$

The ratio term within first bracket in the second term in equation 1.3 can be written as:

$$\frac{\sum_{k=1}^{N_1} \frac{R_{1k}}{\pi_1(Z_{1k})} I(S^c_{1k,obs} = S^c_{1i,obs})\{I(S_{1k} > S_{0j}) - p(S_{1i,obs}, S_{0j})\}}{\sum_{k=1}^{N_1} \frac{R_{1k}}{\pi_1(Z_{1k})} I(S^c_{1k,obs} = S^c_{1i,obs})}$$

$$\approx \frac{1}{q_1(S^c_{1i,obs})} \frac{1}{N_1} \sum_{k=1}^{N_1} \frac{R_{1k}}{\pi_1(Z_{1k})} I(S^c_{1k,obs} = S^c_{1i,obs})\{I(S_{1k} > S_{0j}) - p(S_{1i,obs}, S_{0j})\} \quad (1.5)$$

The ratio term within first bracket in the third term in equation 1.3 can be written as:

$$\frac{\sum_{k=1}^{N_1}\sum_{l=1}^{N_0} \frac{R_{1k}R_{0l}}{\pi_1(Z_{1k})\pi_0(Z_{0l})} I(S^c_{1k,obs} = S^c_{1i,obs}, S^c_{0l,obs} = S^c_{0j,obs})\{I(S_{1k} > S_{0l}) - p(S_{1i,obs}, S_{0j,obs})\}}{\sum_{k=1}^{N_1}\sum_{l=1}^{N_0} \frac{R_{1k}R_{0l}}{\pi_1(Z_{1k})\pi_0(Z_{0l})} I(S^c_{1k,obs} = S^c_{1i,obs}, S^c_{0l,obs} = S^c_{0j,obs})}$$

$$\approx \frac{1}{q_1(S^c_{1i,obs})} \frac{1}{N_1} \sum_{k=1}^{N_1} \left( \frac{R_{1k}}{\pi_1(Z_{1k})} I(S^c_{1k,obs} = S^c_{1i,obs}) E_{S_0|S^c_{0,obs}=S^c_{0j,obs}}[(I(S_{1k} > S_0) - p(S_{1i,obs}, S_{0j,obs}))|S^c_{0,obs} = S^c_{0j,obs}] \right)$$

$$+ \frac{1}{q_0(S^c_{0j,obs})} \frac{1}{N_0} \sum_{l=1}^{N_0} \left( \frac{R_{0l}}{\pi_0(Z_{0l})} I(S^c_{0l,obs} = S^c_{0j,obs}) E_{S_1|S^c_{1,obs}=S^c_{1i,obs}}[(I(S_1 > S_{0l}) - p(S_{1i,obs}, S_{0j,obs}))|S^c_{1,obs} = S^c_{1i,obs}] \right)$$

$$(1.6)$$

Adding equations 1.4, 1.5 and 1.6 and exchanging summations we have:

$$\frac{1}{N_1 N_0} \sum_{i=1}^{N_1} \sum_{j=1}^{N_0} (\hat{\delta}_{ij} - \delta_{ij})$$

$$\approx \frac{1}{N_1} \sum_{k=1}^{N_1} \frac{R_{1k}}{\pi_{1k}} \left( \frac{1}{N_1 N_0} \sum_{i=1}^{N_1} \sum_{j=1}^{N_0} \frac{I(S^c_{1k,obs} = S^c_{1i,obs})}{q_1(S^c_{1i,obs})} \left[ (1-R_{1i})R_{0j}(I(S_{1k} > S_{0j}) - p(S_{1i,obs}, S_{0j})) \right. \right.$$

$$\left. \left. + (1-R_{1i})(1-R_{0j}) E_{S_0|S^c_{0,obs}=S^c_{0j,obs}}[(I(S_{1k} > S_0) - p(S_{1i,obs}, S_{0j,obs}))|S^c_{0,obs} = S^c_{0j,obs}] \right] \right)$$

$$+ \frac{1}{N_0} \sum_{l=1}^{N_0} \frac{R_{0l}}{\pi_{0l}} \left( \frac{1}{N_1 N_0} \sum_{i=1}^{N_1} \sum_{j=1}^{N_0} \frac{I(S^c_{0l,obs} = S^c_{0j,obs})}{q_0(S^c_{0j,obs})} \left[ R_{1i}(1-R_{0j})(I(S_{1i} > S_{0l}) - p(S_{1i}, S_{0j,obs})) \right. \right.$$

$$\left. \left. + (1-R_{1i})(1-R_{0j}) E_{S_1|S^c_{1,obs}=S^c_{1i,obs}}[(I(S_1 > S_{0l}) - p(S_{1i,obs}, S_{0j,obs}))|S^c_{1,obs} = S^c_{1i,obs}] \right] \right)$$

$$(1.7)$$

Using Strong Law of Large Numbers, we approximate the innermost two double-average terms in the above two terms in equation 1.7 with the corresponding expectation. For the first double average we compute the expectation in two pieces. The first piece:

$$E\left[\frac{I(S^c_{1,obs} = S^c_{1k,obs})}{q_1(S^c_{1,obs})}(1-R_1)R_0(I(S_{1k} > S_0) - p(S^c_{1,obs}, S_0))\right]$$
$$= E\left[(1-R_1)R_0(I(S_{1k} > S_0) - p(S^c_{1,obs}, S_0))|S^c_{1,obs} = S^c_{1k,obs}\right]$$
$$= E_{Z_1,Z_0|S^c_{1,obs}=S^c_{1k,obs}}[(1-\pi_1(Z_1))\pi_0(Z_0)E[I(S_{1k} > S_0) - p(S^c_{1,obs}, S_0))|S_{0,obs}, S^c_{1,obs} = S^c_{1k,obs}]|S^c_{1,obs} = S^c_{1k,obs}]$$
$$= E_{Z_1,Z_0|S^c_{1,obs}=S^c_{1k,obs}}[(1-\pi_1(Z_1))\pi_0(Z_0)E[I(S_{1k} > S_0) - p(S^c_{1k,obs}, S_0))|S_{0,obs}, S^c_{1,obs} = S^c_{1k,obs}]|S^c_{1,obs} = S^c_{1k,obs}]$$
$$= E_{Z_1,Z_0|S^c_{1,obs}=S^c_{1k,obs}}[(1-\pi_1(Z_1))\pi_0(Z_0)(p(S_{1k}, S_{0,obs}) - p(S^c_{1k,obs}, S_{0,obs}))|S^c_{1,obs} = S^c_{1k,obs}]$$
$$= E_{Z_1,Z_0|S^c_{1,obs}=S^c_{1k,obs}}[(1-\pi_1(Z_1))\pi_0(Z_0)(p(S_{1k}, S^c_{0,obs}) - p(S^c_{1k,obs}, S^c_{0,obs}))|S^c_{1,obs} = S^c_{1k,obs}]$$

The second piece:

$$E\left[\frac{I(S^c_{1,obs} = S^c_{1k,obs})}{q_1(S^c_{1,obs})}(1-R_1)(1-R_0)E_{S_0|S^c_{0,obs}}[(I(S_{1k} > S_0) - p(S^c_{1,obs}, S^c_{0,obs}))|S^c_{0,obs}]\right]$$
$$= E[(1-R_1)(1-R_0)E_{S_0|S^c_{0,obs}}[(I(S_{1k} > S_0) - p(S^c_{1,obs}, S^c_{0,obs}))|S^c_{0,obs}]|S^c_{1,obs} = S^c_{1k,obs}]$$
$$= E[(1-R_1)(1-R_0)(p(S_{1k}, S^c_{0,obs}) - p(S^c_{1,obs}, S^c_{0,obs}))|S^c_{1,obs} = S^c_{1k,obs}]$$
$$= E[(1-R_1)(1-R_0)(p(S_{1k}, S^c_{0,obs}) - p(S^c_{1k,obs}, S^c_{0,obs}))|S^c_{1,obs} = S^c_{1k,obs}]$$
$$= E_{Z_1,Z_0|S^c_{1,obs}=S^c_{1k,obs}}[(1-\pi_1(Z_1))(1-\pi_0(Z_0))(p(S_{1k}, S^c_{0,obs}) - p(S^c_{1k,obs}, S^c_{0,obs}))|S^c_{1,obs} = S^c_{1k,obs}]$$

We can similarly deal with the second double average in equation 1.7. Hence, equation 1.7 can be further simplified to:

$$\frac{1}{N_1 N_0}\sum_{i=1}^{N_1}\sum_{j=1}^{N_0}(\hat{\delta}_{ij} - \delta_{ij})$$
$$\approx \frac{1}{N_1}\sum_{k=1}^{N_1}\frac{R_{1k}}{\pi_{1k}}E_{Z_1,Z_0|S^c_{1,obs}=S^c_{1k,obs}}[(1-\pi_1(Z_1))(p(S_{1k}, S^c_{0,obs}) - p(S^c_{1k,obs}, S^c_{0,obs}))|S^c_{1,obs} = S^c_{1k,obs}]$$
$$+ \frac{1}{N_0}\sum_{l=1}^{N_0}\frac{R_{0l}}{\pi_{0l}}E_{Z_1,Z_0|S^c_{0,obs}=S^c_{0l,obs}}[(1-\pi_0(Z_0))(p(S^c_{1,obs}, S_{0l}) - p(S^c_{1,obs}, S^c_{0l,obs}))|S^c_{0,obs} = S^c_{0l,obs}]$$
$$= \frac{1}{N_1}\sum_{k=1}^{N_1}\frac{R_{1k}}{\pi_{1k}}\psi_1(S_{1k}) + \frac{1}{N_0}\sum_{l=1}^{N_0}\frac{R_{0l}}{\pi_{0l}}\psi_0(S_{0l})$$

where,

$$\psi_1(S_1) = E_{Z_1, Z_0 | S^c_{1,obs}}[(1 - \pi_1(Z_1))(p(S_1, S^c_{0,obs}) - p(S^c_{1,obs}, S^c_{0,obs})) | S^c_{1,obs}]$$

$$\psi_0(S_0) = E_{Z_1, Z_0 | S^c_{0,obs}}[(1 - \pi_0(Z_0))(p(S^c_{1,obs}, S_0) - p(S^c_{1,obs}, S^c_{0,obs})) | S^c_{0,obs}]$$

**Step 4:** Plugging in the results in **Step 2** and **Step 3** in **Step 1** we have:

$$\begin{aligned}
(\hat{\delta}_{TPS} - \delta_0) &\approx \frac{1}{N_1} \sum_{i=1}^{N_1} R_{1i}(p_0(S_{1i}) - \delta_0) + \frac{1}{N_0} \sum_{j=1}^{N_0} R_{0j}(p_1(S_{0j}) - \delta_0) \\
&+ \frac{1}{N_1} \sum_{i=1}^{N_1} (1 - R_{1i})(p(S_{1i,obs}) - \delta_0) + \frac{1}{N_0} \sum_{i=1}^{N_0} (1 - R_{0j})(p(S_{0j,obs}) - \delta_0) \\
&+ \frac{1}{N_1} \sum_{i=1}^{N_1} \frac{R_{1i}}{\pi_{1i}} \psi_1(S_{1i}) + \frac{1}{N_0} \sum_{j=1}^{N_0} \frac{R_{0j}}{\pi_{0j}} \psi_0(S_{0j}) \\
&= \frac{1}{N_1} \sum_{i=1}^{N_1} \left( R_{1i} p_0(S_{1i}) + (1 - R_{1i}) p(S_{1i,obs}) + \frac{R_{1i}}{\pi_{1i}} \psi_1(S_{1i}) - \delta_0 \right) \\
&+ \frac{1}{N_0} \sum_{j=1}^{N_0} \left( R_{0j} p_1(S_{0j}) + (1 - R_{0j}) p(S_{0j,obs}) + \frac{R_{0j}}{\pi_{0j}} \psi_0(S_{0j}) - \delta_0 \right)
\end{aligned}$$

This gives the influence function representation of $\hat{\delta}_{TPS}$. The derivation of the asymptotic variance formula clearly follows from this step:

$$\begin{aligned}
Var[\hat{\delta}_{TPS}] &\approx \frac{1}{N_1} E\left[ \left( R_1 p_0(S_1) + (1 - R_1) p(S_{1,obs}) + \frac{R_1}{\pi_1(Z_1)} \psi_1(S_1) - \delta_0 \right)^2 \right] \\
&+ \frac{1}{N_0} E\left[ \left( R_0 p_1(S_0) + (1 - R_0) p(S_{0,obs}) + \frac{R_0}{\pi_0(Z_0)} \psi_0(S_0) - \delta_0 \right)^2 \right]
\end{aligned}$$

An estimate of the asymptotic variance can be written as follows:

$$\begin{aligned}
\widehat{Var}[\hat{\delta}_{TPS}] &\approx \frac{1}{N_1^2} \sum_{i=1}^{N_1} \left( R_{1i} \hat{p}_0(S_{1i}) + (1 - R_{1i}) \hat{p}(S_{1i,obs}) + \frac{R_{1i}}{\pi_1(Z_{1i})} \hat{\psi}_1(S_{1i}) - \hat{\delta}_{TPS} \right)^2 \\
&+ \frac{1}{N_0^2} \sum_{j=1}^{N_0} \left( R_{0j} \hat{p}_1(S_{0j}) + (1 - R_{0j}) \hat{p}(S_{0j,obs}) + \frac{R_{0j}}{\pi_0(Z_{0j})} \hat{\psi}_0(S_{0j}) - \hat{\delta}_{TPS} \right)^2
\end{aligned}$$

Here, $\hat{p}_0(S_1)$, $\hat{p}_1(S_0)$, $\hat{p}(S_{1,obs})$, $\hat{p}(S_{0,obs})$, $\hat{\psi}_1(S_1)$ and $\hat{\psi}_0(S_0)$ are estimated empirically using the data from the second phase subjects after adjustment for non-random sampling using sampling weights.

## 2. Variance of the IPW estimator

The IPW estimator of AUC, denoted by $\hat{\delta}_{IPW}$ is obtained by solving the estimating equation: $\sum_{i=1}^{N_1} \sum_{j=1}^{N_0} \frac{R_{1i} R_{0j}}{\pi_1(Z_{1i}) \pi_0(Z_{0j})} (I(S_{1i} > S_{0j}) - \delta) = 0$. Note that:

$$(\hat{\delta}_{IPW} - \delta_0) = \frac{1}{N_1 N_0} \sum_{i=1}^{N_1} \sum_{j=1}^{N_0} \frac{R_{1i} R_{0j}}{\pi_1(Z_{1i}) \pi_0(Z_{0j})} (I(S_{1i} > S_{0j}) - \delta_0) + o_p(1/\sqrt{N})$$

The first term can be expressed as a functional of the corresponding empirical distribution functions: $\theta(F_{S_1}^{(N_1)}, F_{S_0}^{(N_0)}) = \int_{S_1} \int_{S_0} \frac{R_1 R_0}{\pi_1(Z_1) \pi_0(Z_0)} (I(S_1 > S_0) - \delta_0) \, dF_{S_1}^{(N_1)} dF_{S_0}^{(N_0)}$. We also note that $\theta(F_{S_1}, F_{S_0}) = 0$, where $F_{S_1}$ and $F_{S_0}$ are the true distribution functions. A linear von Mises expansion of this term leads to the following: $\int_{S_1} \int_{S_0} \frac{R_1 R_0}{\pi_1(Z_1) \pi_0(Z_0)} (I(S_1 > S_0) - \delta_0) \, dF_{S_1}^{(N_1)} dF_{S_0}^{(N_0)} \approx \frac{1}{N_1} \sum_{i=1}^{N_1} \frac{R_{1i}}{\pi_1(Z_{1i})} (E_{S_0}[I(S_{1i} > S_0)] - \delta_0) + \frac{1}{N_0} \sum_{j=1}^{N_0} \frac{R_{0j}}{\pi_0(Z_{0j})} (E_{S_1}[I(S_1 > S_{0j})] - \delta_0)$. This is the influence function representation of the estimator. The derivation of the asymptotic variance formula clearly follows from this step by noting that each term in the influence function representation has mean 0. The formula and an estimate of the asymptotic variance can be written as follows:

$$Var[\hat{\delta}_{IPW}] \approx \frac{1}{N_1} E\left[\left(\frac{R_1}{\pi_1(Z_1)}\left(p_0(S_1) - \delta_0\right)\right)^2\right] + \frac{1}{N_0} E\left[\left(\frac{R_0}{\pi_0(Z_0)}\left(p_1(S_0) - \delta_0\right)\right)^2\right]$$

$$\hat{Var}[\hat{\delta}_{IPW}] \approx \frac{1}{N_1^2} \sum_{i=1}^{N_1} \left(\frac{R_{1i}}{\pi_1(Z_{1i})}\left(\hat{p}_0(S_{1i}) - \hat{\delta}_{IPW}\right)\right)^2 + \frac{1}{N_0^2} \sum_{j=1}^{N_0} \left(\frac{R_{0j}}{\pi_0(Z_{0j})}\left(\hat{p}_1(S_{0j}) - \hat{\delta}_{IPW}\right)\right)^2$$

$\hat{p}_0(S_1)$ and $\hat{p}_1(S_0)$ are estimated empirically using the data from the second phase subjects after adjustment for non-random sampling using sampling weights.

## 3. Additional details of simulation study

### 3.1 *Computation of sampling weights*

In simple case-control sampling, the sampling weights for the cases are same and is given by the fraction of cases ($\eta$) sampled in the second phase study. The sampling weights for the controls are also same and is the ratio of the number of controls sampled in the second phase and the total number of controls in the cohort.

In stratified case-control sampling, the sampling weights for the cases is $\eta$. We denote by $c_0, \ldots, c_{10}$ the decile cutpoints of the observed riskscore among cases with $c_0 = -\infty$ and $c_{10} = \infty$. The sampling weights for controls in the $j^{th}$ decile is $0.1\eta N_1/N_0 P[c_j < S_{0,obs} \leqslant c_{j+1}]$, $j = 0, \ldots, 10$. The sampling weights are estimated empirically by estimating $c_j$ and $P[c_j < S_{0,obs} \leqslant c_{j+1}]$ from each simulation study.

### 3.2 Additional simulation results

Tables 1 and 2 show the additional simulation results for the cases $f = 0.5$ and $f = 0.2$. The relative efficiency of the proposed two-phase estimator compared to the IPW estimator is smaller for lower values of $f$ due to lower contribution of the variability of the observed risk-score to the variability of the full risk-score. The other performance characteristics are qualitatively similar to the simulation results for $f = 0.75$ reported in the manuscript.

## 4. Additional Details of Data Analysis

### 4.1 Derivation of sampling weights

We derive the sampling weights by fitting a logistic regression model of the indicator of inclusion of a PLCO screening arm participant to the replication study on the case/control status and the following covariates: interaction of age categories ($\leqslant 59$ years, 60-64 years, 65-69 years, $\geqslant 70$ years) and gender (female = 1, male = 0); observed follow-up in 1 year categories; interaction of age categories and smoking pack-year categories and smoking status (current = 1 and former = 2); interaction of age categories and smoking pack-year categories and number of years since stopped smoking; categorical version of year of randomization (levels: 1993-1995, 1995-1997, 1997-1999, 1999-2001) and interactions of all the above covariates with case/control status.

### 4.2  Derivation of relative risk

Table 3 shows the age adjusted relative risks corresponding to the risk factors of three combinations: (i) epidemiologic risk factors, (ii) epidemiological risk factors and C-reactive protein, (iii) epidemiological risk factors and four inflammation biomarkers. The relative risks for the epidemiologic risk factors were reported based on the information from the ever-smokers in the control arm of PLCO Trial (Katki *and others*, 2016). Since information on inflammation biomarkers were not available from those subjects, the relative risks of epidemiologic risk factors for models (ii) and (iii) are not adjusted for inflammation biomarkers.

For the risk factor combination (ii), the relative risk for C-Reactive Protein (CRP) was approximated based on a logistic regression model of case/control status on the quartiles of CRP, categorical age variable (levels: $\leqslant$ 59 years, 60-64 years, 65-69 years, $\geqslant$ 70 years), gender, race, education status, natural logarithm of BMI, smoking pack-year categories, natural logarithm of sum of 1 and smoking quit years, total number of years the subject smoked, indicator of $>1$ pack/day smoked, presence/absence of emphysema/COPD, lung cancer family history, categorical version of year of randomization (levels: 1993-1995, 1995-1997, 1997-1999, 1999-2001). For the risk factor combination (iii), the relative risks of all the four inflammation markers were approximated from another logistic regression model of case/control status on the quartile versions of the four biomarkers and the other variables as specified in the above model. Both these models were based on the discovery study.

Table 1. *Simulation studies evaluating performance of the proposed and the IPW estimators of AUC estimation under two-phase studies. Results are shown under alternative sampling schemes with varying sampling probabilities and a fixed value of $f = 0.5$, the proportion of variability of the full risk-score explained by the partial risk score. The true value of AUC of the underlying model is 0.596. Relative efficiency (RE) of the AUC estimates are reported in reference to standard estimate of AUC obtained using information on all risk factors for all subjects in the cohort. Percent bias in the estimation of standard errors of AUC using the influence function based variance estimate are also reported.*

| Sampling scheme | Bias | | RE (compared to full cohort) | | Percent bias in SE estimate | | Coverage | |
|---|---|---|---|---|---|---|---|---|
| | TPS | IPW | TPS | IPW | TPS | IPW | TPS | IPW |
| Simple case-control sampling | | | | | | | | |
| Sampling fraction of cases 1 | | | | | | | | |
| Case-control ratio 1:1 | $5.85 \times 10^{-5}$ | 0.0002 | 0.61 | 0.49 | 3.33 | 2.99 | 0.94 | 0.94 |
| Sampling fraction of cases 0.5 | | | | | | | | |
| Case-control ratio 1:1 | 0.0004 | 0.0005 | 0.4 | 0.26 | -2.7 | 0 | 0.96 | 0.95 |
| Sampling fraction of cases 0.25 | | | | | | | | |
| Case-control ratio 1:1 | 0.0002 | 0.0005 | 0.19 | 0.13 | 4.63 | 0.76 | 0.95 | 0.95 |
| Stratified case-control sampling | | | | | | | | |
| Sampling fraction of cases 1 | | | | | | | | |
| Case-control ratio 1:1 | 0.0001 | 0.0002 | 0.68 | 0.56 | -1.75 | -4.76 | 0.95 | 0.95 |
| Sampling fraction of cases 0.5 | | | | | | | | |
| Case-control ratio 1:1 | 0.0001 | $2.36 \times 10^{-5}$ | 0.38 | 0.26 | 0 | -1.09 | 0.95 | 0.96 |
| Sampling fraction of cases 0.25 | | | | | | | | |
| Case-control ratio 1:1 | 0.0003 | $5.95 \times 10^{-5}$ | 0.21 | 0.13 | 0 | -0.77 | 0.96 | 0.95 |

Table 2. *Simulation studies evaluating performance of the proposed and the IPW estimators of AUC estimation under two-phase studies. Results are shown under alternative sampling schemes with varying sampling probabilities and a fixed value of $f = 0.2$, proportion of variability of the full risk-score explained by the partial risk-score. The true value of AUC of the underlying model is 0.646. Relative efficiency (RE) of the AUC estimates are reported in reference to standard estimate of AUC obtained using information on all risk factors for all subjects in the cohort. Percent bias in the estimation of standard errors of AUC using the influence function based variance estimate are also reported.*

| Sampling scheme | Bias | | RE (compared to full cohort) | | Percent bias in SE estimate | | Coverage | |
|---|---|---|---|---|---|---|---|---|
| | TPS | IPW | TPS | IPW | TPS | IPW | TPS | IPW |
| Simple case-control sampling | | | | | | | | |
| Sampling fraction of cases 1 | | | | | | | | |
| Case-control ratio 1:1 | $8.66 \times 10^{-5}$ | 0.0001 | 0.59 | 0.55 | 0 | -1.61 | 0.95 | 0.95 |
| Sampling fraction of cases 0.5 | | | | | | | | |
| Case-control ratio 1:1 | -0.0004 | -0.0005 | 0.31 | 0.28 | 0 | -2.3 | 0.95 | 0.96 |
| Sampling fraction of cases 0.25 | | | | | | | | |
| Case-control ratio 1:1 | -0.0002 | -0.0002 | 0.16 | 0.14 | 0 | -1.61 | 0.96 | 0.96 |
| Stratified case-control sampling | | | | | | | | |
| Sampling fraction of cases 1 | | | | | | | | |
| Case-control ratio 1:1 | -0.0005 | -0.0004 | 0.59 | 0.53 | 0 | 0 | 0.95 | 0.96 |
| Sampling fraction of cases 0.5 | | | | | | | | |
| Case-control ratio 1:1 | -0.0005 | -0.0004 | 0.29 | 0.26 | 3.48 | 2.2 | 0.95 | 0.96 |
| Sampling fraction of cases 0.25 | | | | | | | | |
| Case-control ratio 1:1 | -0.0003 | -0.0001 | 0.16 | 0.14 | 0.86 | -1.61 | 0.95 | 0.96 |

Table 3. *Relative risks of the risk factors estimated using information from the control arm of PLCO Trial and the discovery study. The relative risks for the epidemiologic risk factors (i.e., demographics, smoking, lung disease, family history) are obtained from Katki and others (2016) based on the ever smokers in the control arm of the PLCO trial. The relative risks for the inflammation markers are estimated, after adjustment of other covariates, based on the ever smokers in the discovery study, a nested case-control study within the screening arm of the PLCO trial. "Education" was coded as: less than grade 12 = 1, high school graduegeate = 2, post high school but no college = 3, some college = 4, bachelor's degree = 5, graduate school = 6. "Quit years", i.e., number of years since stopped smoking cigarettes was added as natural logarithm of 1 plus quit-years. Lung cancer family history was defined as the number of first degree relatives (siblings/parents/children) with history of lung cancer*

| | | Relative risk | | |
|---|---|---|---|---|
| Risk factor | Coding | Epidemiologic risk factors only | Epidemiologic risk factors + CRP | Epidemiologic risk factors + 4 inflammation markers |
| **Demographics** | | | | |
| Female sex | Binary | 0.92 | 0.92 | 0.92 |
| Race | Categorical | | | |
| | White, non-Hispanic | 1 [Reference] | 1 [Reference] | 1 [Reference] |
| | Black, non-Hispanic | 1.24 | 1.24 | 1.24 |
| | Hispanic | 0.65 | 0.65 | 0.65 |
| | Asian or other | 0.67 | 0.67 | 0.67 |
| Education | Trend | 0.93 | 0.93 | 0.93 |
| BMI | Log-term | 0.49 | 0.49 | 0.49 |
| **Smoking** | | | | |
| Pack-years | Categorical | | | |
| | 0-29.9 | 1 [Reference] | 1 [Reference] | 1 [Reference] |
| | 30-39.9 | 1.63 | 1.63 | 1.63 |
| | 40-49.9 | 1.76 | 1.76 | 1.76 |
| | >50 | 2.05 | 2.05 | 2.05 |
| Quit years | Log-term | 0.73 | 0.73 | 0.73 |
| Years smoked | Linear | 1.02 | 1.02 | 1.02 |
| >1 pack/day | Binary | 1.36 | 1.36 | 1.36 |
| **Lung disease** | | | | |
| Emphysema/COPD | Binary | 1.76 | 1.76 | 1.76 |

Table 3 (continued).

| Risk factor | Coding | Relative risk | | |
| --- | --- | --- | --- | --- |
| | | Epidemiologic risk factors only | Epidemiologic risk factors + CRP | Epidemiologic risk factors + 4 inflammation markers |
| **Family history** | | | | |
| Lung cancer family history | Trend | 1.52 | 1.52 | 1.52 |
| **Inflammation biomarkers** | | | | |
| CRP | Categorical | | | |
| | Quartile 1 | | 1[Reference] | 1 [Reference] |
| | Quartile 2 | | 1.52 | 1.42 |
| | Quartile 3 | | 1.45 | 1.30 |
| | Quartile 4 | | 2.20 | 1.82 |
| SAA | | | | |
| | Quartile 1 | | | 1[Reference] |
| | Quartile 2 | | | 1.09 |
| | Quartile 3 | | | 1.32 |
| | Quartile 4 | | | 1.35 |
| sTNFRII | | | | |
| | Quartile 1 | | | 1[Reference] |
| | Quartile 2 | | | 0.98 |
| | Quartile 3 | | | 1.13 |
| | Quartile 4 | | | 0.90 |
| CXCL9/MIG | | | | |
| | Quartile 1 | | | 1 [Reference] |
| | Quartile 2 | | | 0.73 |
| | Quartile 3 | | | 0.99 |
| | Quartile 4 | | | 1.03 |